\pdfoutput=1
\documentclass[letterpaper, 10 pt, conference]{ieeeconf}  

\usepackage{graphicx}
\usepackage{cite}
\usepackage{amsmath,amssymb,amsfonts}
\usepackage{algorithmic}
\usepackage{textcomp}
\usepackage{xcolor}
\usepackage{tikz}

  \newcommand\copyrighttext{%
  \footnotesize \textcopyright 2021 IEEE. Personal use of this material is permitted.  Permission from IEEE must be obtained for all other uses, in any current or future media, including reprinting/republishing this material for advertising or promotional purposes, creating new collective works, for resale or redistribution to servers or lists, or reuse of any copyrighted component of this work in other works. DOI: 10.1109/ICAR53236.2021.9659317}
\newcommand\copyrightnotice{%
\begin{tikzpicture}[remember picture,overlay]
\node[anchor=south,yshift=10pt] at (current page.south) {\fbox{\parbox{\dimexpr\textwidth-\fboxsep-\fboxrule\relax}{\copyrighttext}}};
\end{tikzpicture}%
}

\IEEEoverridecommandlockouts                              

\title{\LARGE \bf
A High-Level Track Fusion Scheme for Circular Quantities*
}

\author{Sören Kohnert$^{1}$ and Reinhard Stolle$^{1}$
\thanks{*This work is supported by Siemens Mobility GmbH.}
\thanks{$^{1}$The authors are with Faculty of Electrical Engineering,
        Augsburg University of Applied Science, 86161 Augsburg, Germany.
        {\tt\small soeren.kohnert@hs-augsburg.de}}%
}

\begin{document}

\maketitle
\copyrightnotice
\thispagestyle{empty}
\pagestyle{empty}


\begin{abstract}
As sensors get more and more integrated, signal processing functions, like tracking, are performed closer to the sensor. Consequently, high level fusion is on the rise. Presented here is a high level fusion scheme incorporating not only linear, but circular quantities as well. Monte Carlo experiments are used to verify our novel fusion operators that work as a weighted average for the Wrapped Normal and the von-Mises distribution. To further verify the new fusion operators, we implemented a full track level fusion scheme and tested it by fusing the measurements of two RADAR sensors.
\end{abstract}



\section{Introduction}
Perception is a key component of automated vehicle stacks. Typically, to reach a high Automotive Safety Integrity Level (ASIL), multiple independent sensors with low ASILs are combined using ASIL decomposition \cite{iso26262}. Data from different sensors need to get fused. The data mostly consists of linear quantities like velocity and position estimates, but some circular quantities like the heading of objects need special care. The nature of circular values stands out e.g. due to ambiguities.
Whereas the trend goes to more and more integrated sensors, the fusion level shifts from low and feature level fusions to high level fusions such as track-to-track (T2T) level fusions \cite{aeberhard_20}. 
One possibility to do high level fusion is to combine the data maximum likelihood wise. High level fusion schemes exist for linear quantities \cite{Houenou_2012}, but they neglect circular quantities so far. Literature on the application of circular quantities \cite{Borradaile2003} mainly focuses on earth science topics. To our knowledge there is no dedicated high level fusion scheme that incorporates circular quantities. Stienne et al. \cite{Stienne2014} follow a similar approach as this contribution, but take a different fusion operator for the dispersion.
Whereas statistical analysis of orientation data is complex, most engineers only require a few simple estimates of population parameters and the corresponding fusion operators.
This paper explores three issues:
\begin{itemize}
 \item It provides an understanding of how to handle circular quantities and their variance and dispersion measures.
 \item It presents evidence based fusion operators based on the weighted average and mean.
 \item It presents a T2T fusion scheme incorporating both, circular and linear quantities.
\end{itemize}

In section II, two circular distributions are introduced and multiple measures for mean and dispersion discussed. In section III, fusion operators are derived and then verified in section IV using Monte Carlo (MC) experiments. Additionally, it is shown why we think this approach is superior to the approach by Stienne et al.\cite{Stienne2014}. Section V focuses on the track-to-track fusion scheme that incorporates circular quantities, and on a real world test of the proposed scheme in an infrastructure-based perception application. Chapter VI summarizes and discusses the main results.

\section{Circular Statistics}
Circular statistics have some peculiarities that deserve special attention. The \textit{crossover} problem is probably the most prominent one. If two orientations have azimuth $350\deg$ and $10\deg$, it is easy to see that their mean is $0\deg$ and not $180\deg$. The following section introduces two common circular distributions and concepts to deal with these pitfalls properly. 

\subsection{Circular distributions}
The two following symmetric circular distributions are most common. They are both closely related to the Normal distribution $\mathcal{N}$ on the real line and diverge by a few percentages only, and they are often used interchangeably as an approximation of one another~\cite{Jammalamadaka2001}. The location parameter for both distributions is $\mu$.
The von-Mises distribution $\mathcal{VM}$ is also known as the circular Normal distribution inter alia it is the maximum entropy distribution for the circular case \cite{Jammalamadaka2001}. $\kappa$ is the concentration parameter. It can be seen as a reciprocal of a dispersion measure. The smaller the concentration parameter, the more evenly distributed is the distribution. The von-Mises distribution can be described as
\begin{equation}
f_{\mathcal{VM}}(\theta\mid\mu,\kappa)=\frac{1}{2\pi I_0(\kappa)}\exp \left[ {\kappa\cos(\theta-\mu)} \right],
\label{eq_vonMises}
\end{equation}
where $I_n$ is the modified Bessel function of order $n$.
The Wrapped Normal distribution $\mathcal{WN}$ is the linear Normal distribution $\mathcal{N}$ wrapped around a circle. Like the normal distribution, it has the variance $\sigma^2$ as dispersion parameter. The Wrapped Normal distribution is defined as 
\begin{equation}
f_{\mathcal{WN}}(\theta\mid\mu,\sigma)=\frac{1}{\sigma \sqrt{2\pi}} \sum^{\infty}_{k=-\infty} \exp \left[\frac{-(\theta - \mu + 2\pi k)^2}{2 \sigma^2} \right].
\label{eq_wrappedNormal}
\end{equation}
For high values of the concentration parameter $\kappa$ and small values of the variance $\sigma^2$, both distributions can be seen as approximations of the Normal distribution $\mathcal{N}$ since only a fraction of the samples fall into the ambiguous area $ | \theta | > \pi$.

\subsection{Circular Mean Resultant Length and Orientation}
Among others, Fisher \cite{Fisher1987} described the crossover problem for circular data. Treating the values as vectors overcomes the problem and gives the resultant vector $R$. $R$ is the line joining the start and endpoints of the sample vectors when they are stacked onto each other under consideration of their orientation. Fig.~\ref{fig_resultantVector} illustrates this approach. 
It is possible to either present the angles in the complex plane using unit vectors with the argument of the angular samples, or to decompose the resulting vector R in the two components:
\begin{align}
C =  \sum_{i=1}^{n}\cos{\theta_i} && \mathrm{and} && S =  \sum_{i=1}^{n}\sin{\theta_i}, 
\label{eq_resultingVectorC}
\end{align}
where $\theta_i$ are the samples and  $n$ is the number of samples. $C$ and $S$ are the coordinate components of vector $R$. For example, $S$ is the mean length of the projection on the \mbox{y-axis}. It is not necessary to know the orientation distribution of the population or to assume any theoretical model; the method is universally valid \cite{Borradaile2003}. The resulting vector $R$ is a direct measure of mean orientation and dispersion:

 \begin{figure}[ht]
\input{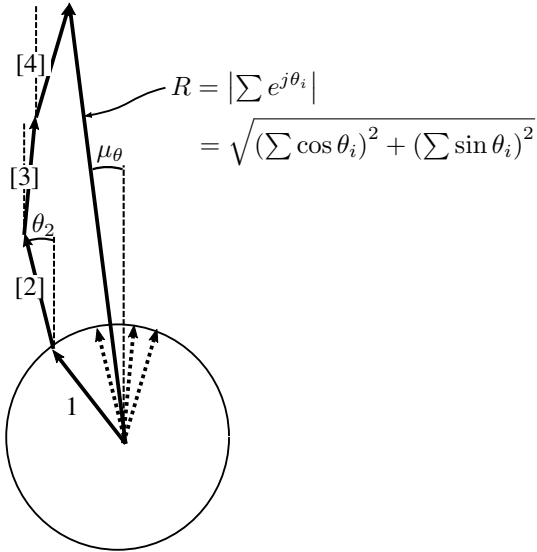}
\caption{The samples are represented by their orientation and have a length of one. The samples are stacked on top of each other under consideration of their orientation. The resultant vector $R$ serves as measure of mean orientation and dispersion. Source: Adapted from \cite{Borradaile2003}. }
\label{fig_resultantVector}
\end{figure}


\subsubsection{Mean Orientation}
The mean orientation of the samples is the orientation of the resulting vector $R$
\begin{equation}
\mu_{\theta}  = \arctan_{2D} ( S , C ),
\label{eq_meanOrientation}
\end{equation}
where $\arctan_{2D}$ is the four-quadrant inverse tangent. 

\subsubsection{Dispersion}
 From the resultant vector $R$, the first circular moment can be derived by
  \begin{equation}
\Bar{R} =  (1/n \cdot C) + (1/n \cdot S) .
\label{eq_firstCircularMoment}
\end{equation}
Thus, $\bar{R}$ is the mean vector.
Subsequently, the estimate for the mean resultant length 
 \begin{equation}
||\Bar{R}|| = \sqrt{ (1/n \cdot C)^2 + (1/n \cdot S)^2 } = \frac{1}{n}\left|\sum e^{j\theta_i}\right|
\label{eq_dispersionOrientation}
\end{equation}
can be obtained as a form of dispersion measure.
The closer the mean resultant length $||\Bar{R}||$  matches the unit vector length, the smaller the dispersion of the distribution. This is how concentration parameters behave. The concentration parameter is e.g. defined for $\mathcal{N}$ as $1/\sigma^2$. 
 The mean resultant length $R$ is the natural dispersion measure for circular distributions but has, in contrast to variance definitions, not the same (nor squared) (pseudo) unit as the quantity it describes. 

\subsection{Variance}
Circular variance is defined as 
\begin{equation}
V = 1-||\Bar{R}||, 
\label{eq_circVar}
\end{equation}
with $0 \le V \le 1$ \cite{Borradaile2003}. In contrast to the variance $\sigma^2$ the circular variance $V$ has an upper limit. The unit of the circular variance is a dimensionless length, not an angle.
Both properties make it difficult to compare this measure to the variance definition for linear values.
A definition depending on $R$ is beneficial, if further statistical analysis is done, but
most engineers are familiar with Gaussian distributions on the real line and consequently are familiar with linear variances. So it makes sense to have an equivalent measure in the circular domain.
\subsection{An Unbiased Sample Variance Definition Inspired by the Definition on the Real Line}
A sample variance can be defined equivalent to the variance on the real line.
The minimum distance between two points on a circle \cite{Farrugia2009} is

\begin{equation}
\Delta(\theta_i, \alpha) = 2 \cdot \arctan \left( \tan \left( 0.5 \cdot \left( \theta_i - \alpha \right) \right) \right) .
\label{eq_minDistOnCircle}
\end{equation}

Stupavsky and Symons \cite{Stupavsky1982} suggest using an estimate by analogy with the Normal distribution's unbiased sample variance:
\begin{equation}
\hat{\sigma}^2_{\mathrm{S\&S}} = \frac{1}{n-1} \cdot 
\sum_{i=1}^{n} \left( \Delta(\theta_i, \alpha)^2 - \left( \frac{1}{n} \cdot \sum_{i=1}^{n} \Delta(\theta_i, \alpha)  \right) ^2 \right).
\label{eq_Stupavsky}
\end{equation}
The use of this formula might be a good idea if the underlying distribution of the samples is unknown.

\subsection{Using the Unbiased Estimate of the Squared Mean Resultant Length $\bar{R}^2$}
Kutil \cite{Kutil_2012} showed another approach by taking the second circular moment

\begin{equation}
\Bar{R}^2 = (1/n \cdot S)^2 + (1/n \cdot C)^2 
\label{eq_secondCircularMoment}
\end{equation}
and calculating its expected value 
\begin{equation}
\left\langle \overline{R}^2\right\rangle = \frac{1}{n}+\frac{n-1}{n} a,
\label{eq_RestUnbiase}
\end{equation}
where $a$ is $e^{-\sigma^2}$ for $\mathcal{WN}$ and $I_1(\kappa)^2 / I_0(\kappa)^2$ for $\mathcal{VM}$. $I_n$ is the modified Bessel function of order $n$.
From that, one can derive the expected circular variance for the Wrapped Normal distributions as
\begin{equation}
\hat{\sigma}^2_{\mathcal{WN}} = -\ln \left( \frac{n}{n-1} \cdot \left( \bar{R}^2 - \frac{1}{n} \right) \right)
\label{eq_secondCircMomentVariance}
\end{equation}
which is equal to $\sigma^2$ if $\bar{R}^2$ is unbiased.

There is no closed-form expression for the inverse of the Bessel function, as a result solving \eqref{eq_RestUnbiase} for $\kappa$ is not straight forward. There are several approximations available, see e.g. \cite{fisher_1993}. Banerjee \cite{banerjee05a} gave a simple approximation for the problem using the first \eqref{eq_firstCircularMoment} and second \eqref{eq_secondCircularMoment} circular moment 
\begin{equation}
\hat{\kappa} = \frac{\bar{R}(p-\bar{R}^2)}{1-\bar{R}^2},
\label{eq_Banerjee05}
\end{equation}
where $p=2$ in the case of the 2-variate $\mathcal{VM}$.
\begin{figure}[ht]
\centerline{\includegraphics[trim=0 0 0 15, clip,width=80mm]{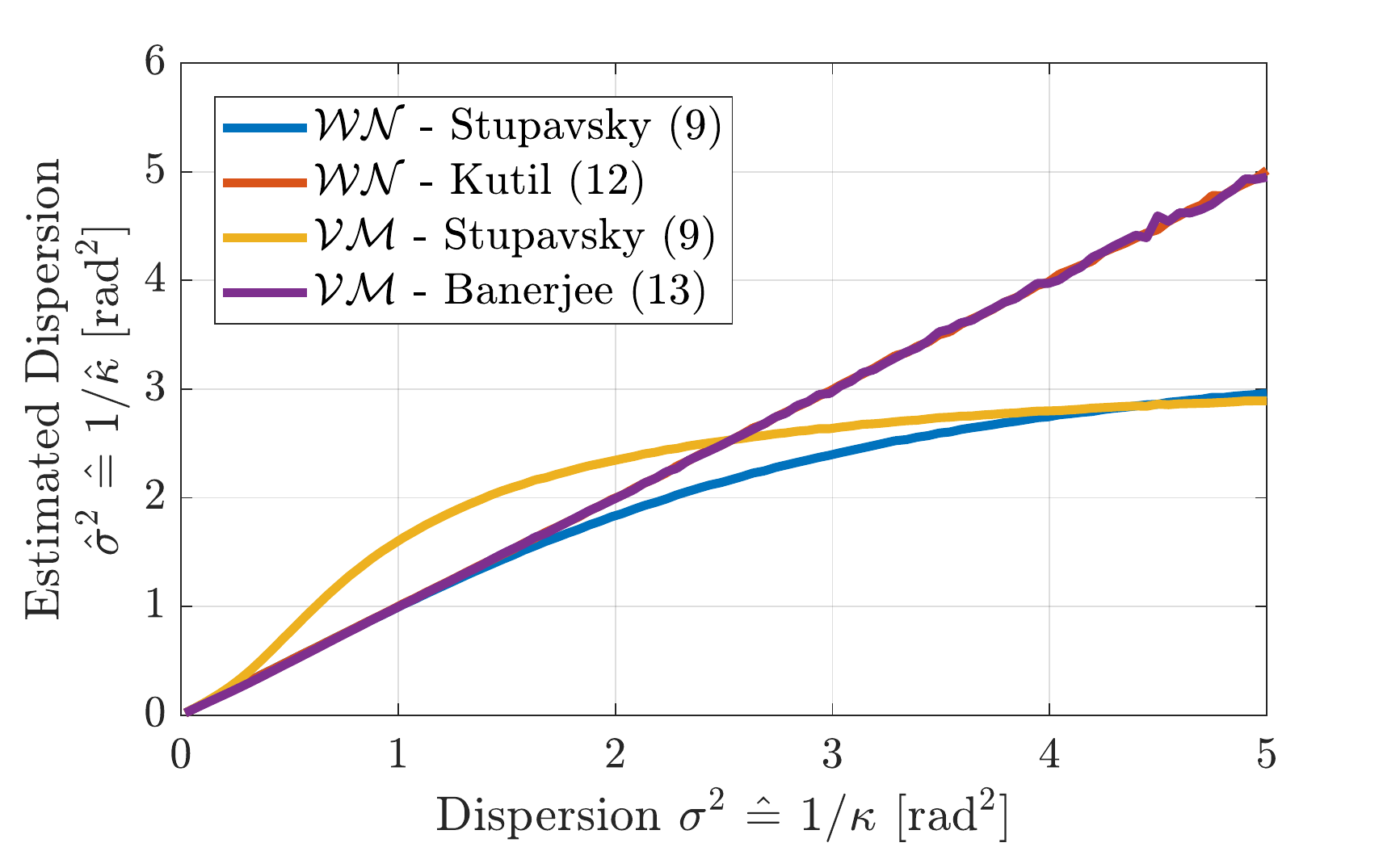}}
\caption{Comparison of the different formulas to derive the sample variance for the $\mathcal{WN}$ and $\mathcal{VM}$ distributions. The variance definition using the second circular moment \eqref{eq_secondCircMomentVariance} works best for $\mathcal{WN}$, whereas \eqref{eq_Banerjee05} works well for $\mathcal{VM}$ distribution.}
\label{fig_sampleVarianc}
\end{figure}
To compare the different measures of dispersion, a Monte Carlo experiment was done using $10^5$ samples. The results are shown in Fig.~\ref{fig_sampleVarianc}. The $\mathcal{VM}$ and $\mathcal{WN}$ were sampled for different values of $\kappa$ and $\sigma^2$, respectively. Their sampled variance was estimated using the formula of Stupavky and Symons \eqref{eq_Stupavsky} for both distributions, the second circular moment approach \eqref{eq_secondCircMomentVariance} for the $\mathcal{WN}$, and Banerjee's formula for the $\mathcal{VM}$ distribution. It becomes apparent that Stupavskys formula is accurate for small dispersions of the $\mathcal{WN}$, but settles against $\pi^2 / \sqrt{3}$ which is the variance expected for a symmetric rectangular distribution from $-\pi$ to $\pi$ \cite{gum}. For this reason, we will stick to the definitions by Banerjee \eqref{eq_Banerjee05} and Kutil \eqref{eq_secondCircMomentVariance} for the rest of the paper.

\section{Fusion Operator}
We propose possible fusion operators for the $\mathcal{VM}$ and $\mathcal{WN}$ and verify their results using MC experiments. The starting point are two sensors that deliver estimates of the circular mean as well as estimates of the dispersion to the fusion unit. Both sensors are assumed to be statistically independent.

\subsection{Maximum Likelihood Estimator for Mean with Known Dispersion}
A weighted average (wavrg.) is the maximum likelihood estimation for the Normal distribution in the linear case \cite{Koch2013}. Stienne et al. \cite{Stienne2014} derived the maximum likelihood estimate for the $\mathcal{VM}$ distribution as 

\begin{equation}
\mu_{\theta, \mathrm{fused, wavrg.}} = \arctan_{2D}(S_f,C_f),
\label{eq_mlMu}
\end{equation}
with
\begin{equation}
C_f =   \sum_{i=1}^{n}\kappa_i \cdot \cos{\theta_i}  \hat{=} \sum_{i=1}^{n} 1/\sigma_i^2 \cdot \cos{\theta_i}
\label{eq_resultingVectorC_weighted}
\end{equation}
and
\begin{equation}
S_f =  \sum_{i=1}^{n}\kappa_i \cdot \sin{\theta_i} \hat{=} \sum_{i=1}^{n} 1/\sigma_i^2 \cdot\sin{\theta_i}.
\label{eq_resultingVectorS_weighted}
\end{equation}
This is quite intuitive, since we weight the lengths of the single components by their concentration parameter $\kappa$ (c.f. Fig.~\ref{fig_resultantVector}). This is analogous to the wavrg. for Normal distributions. We will use the same result later on for the wrapped Normal distribution (using the reciprocal of the variance $\sigma^2$) and will see that it makes a good estimator as well.

\subsection{Estimators for Dispersion and Concentration}
Assuming statistically independent sensors, a formula for the resulting variance in the case of quantities on the real line can be derived.
We will use the same formula for the circular case, since the fusion of the variance is independent from the position of the mean. 
Therefore, not suffering from e.g. the crossover problem.
We assume the estimators
\begin{equation}
\frac{1}{\sigma^2_\mathrm{fused, wavrg.}} = \sum_{i=1}^n \frac{1}{\sigma^2_i}
\label{eq_sigmaFusedWeighted}
\end{equation}
and 
\begin{equation}
 \kappa_\mathrm{fused, wavrg.} = \sum_{i=1}^n \kappa_i
\label{eq_kappaFusedWeighted}
\end{equation}
for the  $\mathcal{VM}$ and $\mathcal{WN}$, respectively.
If one takes the mean  of the measurements \eqref{eq_meanOrientation}, we suggest to inspire for the corresponding variance and dispersion by the Normal distribution $\mathcal{N}$  as well:
\begin{equation}
{\sigma^2_\mathrm{fused, mean}} = \frac{\sum_{i=1}^n \sigma^2_i}{n^2}
\label{eq_sigmaFused}
\end{equation}
and
\begin{equation}
\frac{1}{\kappa_\mathrm{fused, mean}} = \frac{\sum_{i=1}^n 1 / \kappa_i}{n^2}.
\label{eq_kappaFused}
\end{equation}
Taking the mean corresponds to assuming the same variance and dispersion for both measurements, respectively. I.e., applying equal weights on both measurements during fusion.

\section{Monte Carlo Experiments}
To verify the estimators for the variance of the fused measurements, an MC experiment was done using $10^5$ samples. The source code used to do the MC experiments is published under a MIT license.\footnote{github.com/soerenkoh/CircularFusionOperator} The measurements of two sensors are fused together as follows. First, the wavrg. approach is applied, using \eqref{eq_mlMu} for the fused mean and \eqref{eq_sigmaFusedWeighted} and \eqref{eq_kappaFusedWeighted} for the resulting dispersion and concentration, respectively. Second, a simple averaging (mean) is performed with the fused estimate for the mean \eqref{eq_meanOrientation} and the resulting dispersion \eqref{eq_sigmaFused} and concentration \eqref{eq_kappaFused}. 
\begin{figure}[ht]
\centerline{\includegraphics[trim=0 0 0 0, clip,width=80mm]{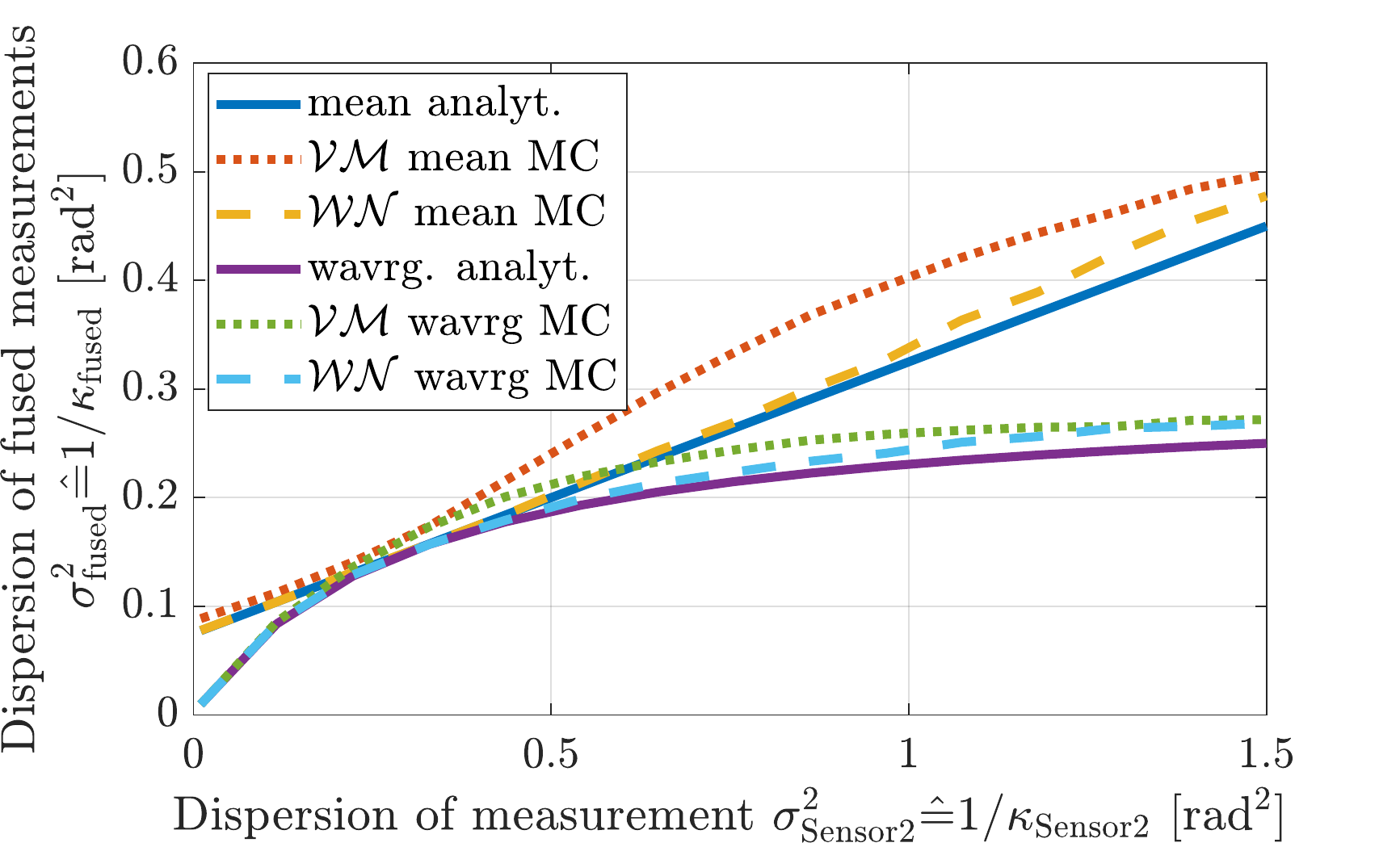}}
\caption{
 As the second sensor's measurement dispersion approaches 1.5, the weighted estimate approaches 0.3: The variance of sensor~1's measurement. This is to be expected, since no new valuable information gets added through the second sensor anymore. This is true for both distributions.} 
\label{fig_soerenfusion}
\end{figure}
In Fig.~\ref{fig_soerenfusion}, results are shown for the dispersion of the estimated fused measurement, overlaid with the MC simulation. Sensor 1 provides measurements with a fixed variance of $\sigma_\mathrm{Sensor1}^2 = 0.3\,\mathrm{rad}^2$ whereas, the variance of sensor~2 is made variable. The fusion result is expected to have zero variance if one sensor has zero variance for the wavrg. case. In the case of both sensors providing measurements with equal variance, the variance of the wavrg. should not differ from the variance of the mean. In case of a high variance of one sensor, the fused variance should converge to the variance of the measurement with the lower variance.
This is the case for both distributions. We see that the Monte Carlo result for the $\mathcal{WN}$ distribution follows the result of our fusion operator quite well, whereas the $\mathcal{VM}$ distribution diverges, especially for high variances. The latter will have little impact on practical applications. The lower the variance (the higher the dispersion) of the measurements, the more similar the distributions get to the $\mathcal{N}$ distribution, and the closer the proposed fusion operator follows the MC result. 
The fusion operator is expected to be more precise for the $\mathcal{WN}$ distribution since the $\mathcal{WN}$ distribution, in contrast to the $\mathcal{VM}$, is closed under convolution \cite{Jammalamadaka2001}. The fusion proposed here can be expressed as a convolution of the distributions of the measurements. Consequently, the result of a fusion of measurements that follow a $\mathcal{VM}$ does not follow a $\mathcal{VM}$ anymore.

\begin{figure}[ht]
\centerline{\includegraphics[trim=0 0 0 0, clip,width=80mm]{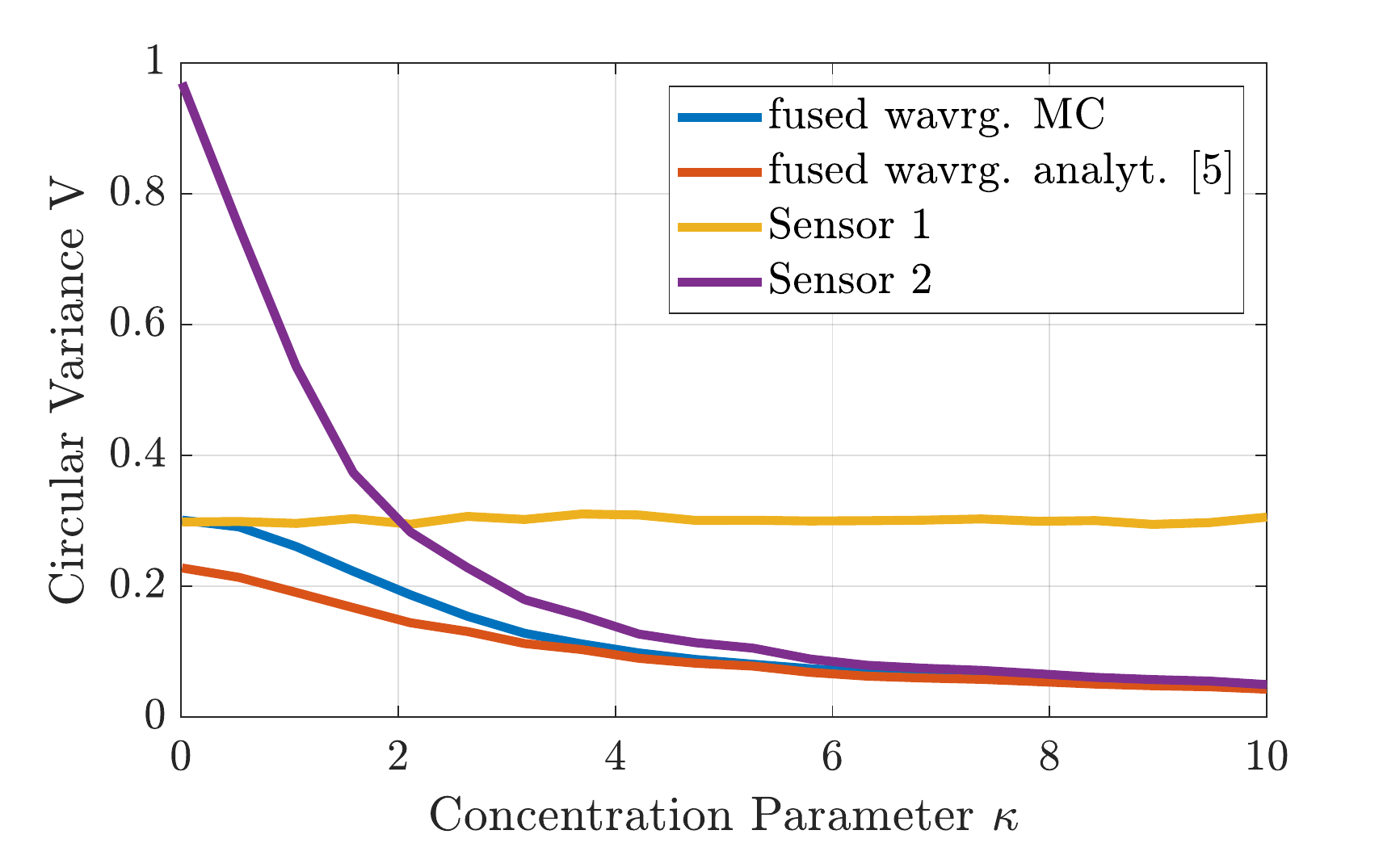}}
\caption{Fusion based on circular variance as a fusion parameter. Result is shown for a $\mathcal{VM}$ with $\kappa_{Sensor 1} = 0.5$. This is an edited reproduction of \cite{Stienne2014}. When one measurement has high dispersion, the circular variance of the fused measurements does not converge against the dispersion of the second measurement with lower circular variance.}
\label{fig_stienneFusion}
\end{figure}

Stienne et al. \cite{Stienne2014} propose to use the circular variance $V$ as a fusion parameter.  Using
\begin{equation}
\frac{1}{V_\mathrm{fused, wavrg.}} = \sum_{i=1}^n \frac{1}{V_i}
\label{eq_circVarFused}
\end{equation}
as an estimate for the variance of the expected values after fusion. Their definition is analogous to the expression presented here \eqref{eq_sigmaFusedWeighted} and \eqref{eq_kappaFusedWeighted}, but uses $V$ instead of $\sigma^2$ and $\kappa$, respectively. Our simulation indicates that it is not as well suited as the estimation based on the variance $\sigma^2$ or on the concentration parameter $\kappa$. The condition
\begin{equation}
{V_\mathrm{fused, wavrg.}} < \min{V_{i}} 
\label{eq_circVarFusedmin}
\end{equation}
is true, but the fused circular variance $V_\mathrm{fused, wavrg.}$ does not approach the smaller value of the circular variance $V_{i}$ of the measurements when the dispersion is increasing since the circular variance is limited to 1 (cf. \eqref{eq_circVar}). The greater the value for the smaller $V$, the greater this error gets. The results of the Monte Carlo experiment shown in  Fig.~\ref{fig_stienneFusion} illustrate the finding. A repetitive fusion of the measurements would lead to a lower circular variance $V_\mathrm{fused, wavrg.}$ each time.


\section{Track-to-Track Fusion}
To see the effect of the circular fusion on real measurement data, we implemented a T2T fusion scheme based on the linear fusion scheme of \cite{Houenou_2012}, but extended it with the circular fusion operators presented above. The new fusion scheme provides a way to fuse circular values like the heading of objects. The fusion scheme is tested using an infrastructure-based perception application. 

\subsection{Fusion Scheme}
\begin{figure}[ht]
\centerline{\includegraphics[trim=0 0 0 96, clip,width=75mm]{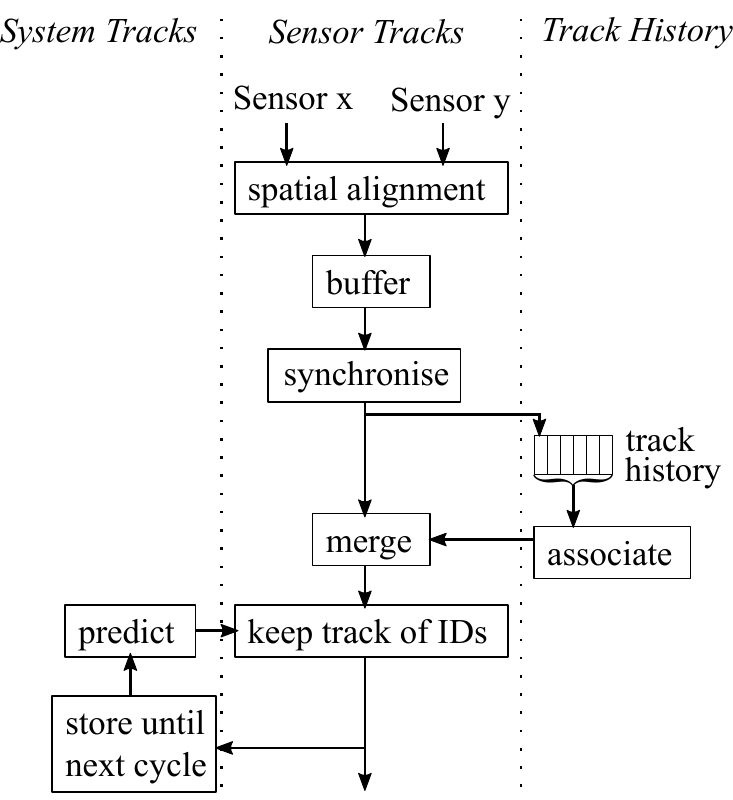}}
\caption{Track-to-Track Fusion scheme based on \cite{Houenou_2012}. The data flow is from top to bottom. The concept of track history is used to resolve associations incorporating the history of the tracks. System tracks are used to ensure persistent track IDs over multiple fusion cycles.}
\label{fig_t2tfusion}
\end{figure}

The fusion scheme used here is shown in Fig.~\ref{fig_t2tfusion}. The tracks generated by multiple sensors are first aligned spatially. Next, to overcome the problem of out-of-sequence measurements, the tracks are buffered. The fusion is done at a fixed fusion rate.
A constant velocity model is used for temporal alignment of the buffered sensor tracks. The temporally aligned tracks are saved in the track history. The track history is a fixed size buffer, e.g. for the tracks of the last six fusion cycles. This method is suitable for situations in which objects  are close to each other for one moment, but come from different directions.

The association is done using a Global Nearest Neighbour (GNN) approach with a variant of the Mahalanobis distance as distance measure. The associated log likelihood distance \cite{altendorfer_2015} limits the Mahalanobis distance so that tracks with a very low variance do not "steal" an association from a track with high variance. If it is desired to incorporate the circular values into the association, it is possible to use the expression for the minimum distance on the circle \eqref{eq_minDistOnCircle} to calculate the distance between the circular states. Suppose that the circular dispersion measures $\sigma^2$ and $1/\kappa$ can be used analogous to the linear ones in the covariance matrix of the states. 

In the merging step, the states of the tracks are fused using the maximum likelihood estimate for the mean angle \eqref{eq_mlMu} and its estimate for the fused variance \eqref{eq_sigmaFusedWeighted} and dispersion \eqref{eq_kappaFusedWeighted}, respectively.  Keeping track of the fused track IDs is the last step. The fused tracks get associated to (temporally aligned) system tracks of the last fusion cycle, making it possible to have coherent track IDs over all fusion cycles. GNN is used here as well to find a solution to the association problem.

\subsection{Measurements}
Even though the fusion scheme can handle multiple objects, we concentrate on a single object for the sake of clarity in this paper. Due to the lack of ground truth data, we can not show the true position and orientation of the object. Nevertheless, the conclusion can be drawn that the wavrg. works in a real world implementation, if the data is free of any crossover problem and the results look reasonable. 

\begin{figure}[ht]
\centerline{\includegraphics[trim=0 80 0 98, clip,width=80mm]{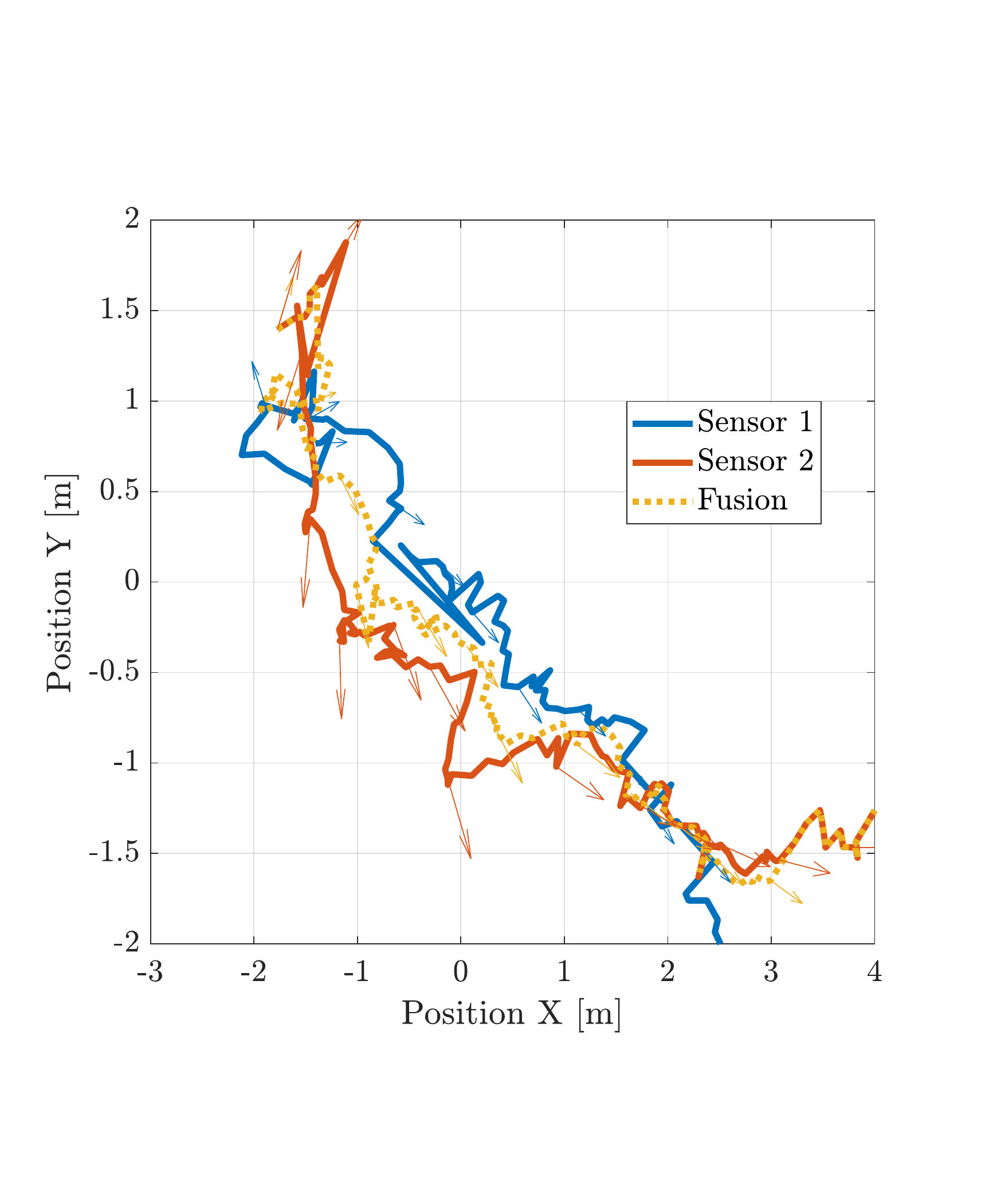}}
\caption{Position of pedestrian as seen by sensor~1 (blue) and sensor~2 (red) as well as the result of the fusion of both estimates (yellow dotted line).}
\label{fig_t2tPosition}
\end{figure}

The path of the object is shown in Fig.~\ref{fig_t2tPosition}. Sensors and fusion run at a fixed rate of $18\,\mathrm{Hz}$.
The object is a pedestrian, first standing at approx. (-1,1) and later walking to the south-east. Fig.~\ref{fig_t2tPosition} shows a street section that is monitored by two independent radar sensors. Sensor~1 is located at position $(-8.2, 15.8)$ and sensor~2 at position $(-18.2, -2.8)$.  To show the robustness of the proposed fusion operators, a fusion of tracks with significantly different heading estimates in the beginning was chosen. Especially in the beginning when the pedestrian is standing, the velocity based heading estimation of the single sensors is far off. Fig.~\ref{fig_t2tPosition} shows the tracks of both sensors as well as the result of the (linear) T2T fusion, basically doing a wavrg. between both tracks. The small arrows indicate the estimate of the heading at some places along the way. After the track of sensor~1 disappears because the pedestrian walks out of it's field-of-view (FoV) at $(2.5, -2)$, the fused track follows exactly the track of sensor~2. 
\begin{figure}[ht]
\centerline{\includegraphics[trim=0 0 0 19, clip,width=80mm]{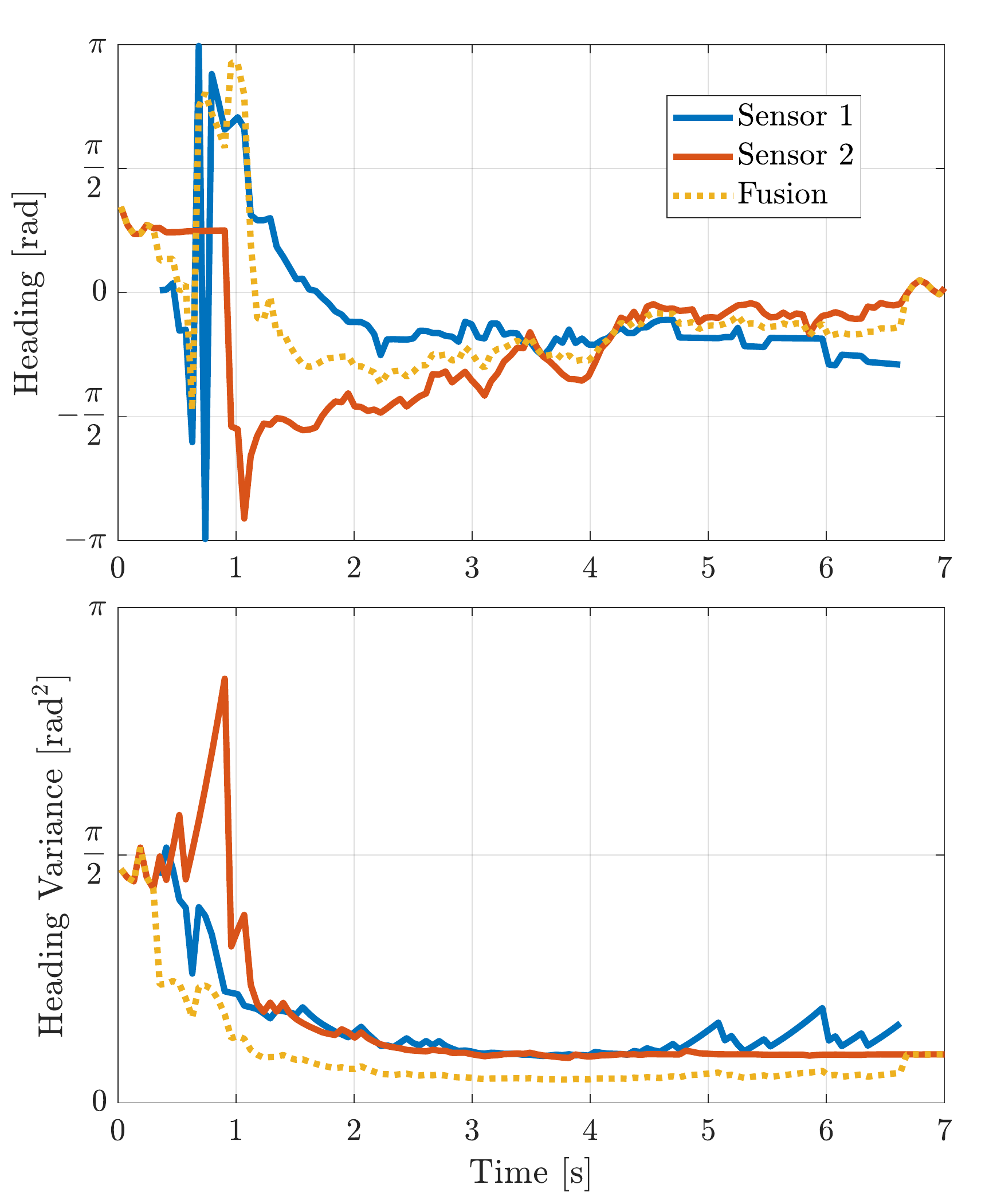}}
\caption{Heading and it's variance over time of the pedestrian's walk from Fig.~\ref{fig_t2tPosition}. The sensors derive the heading and it's covariance from the linearisation of the velocity. Since the pedestrian was standing still at first and then started moving, the sensors get more sure about the object's orientation as time passes. The estimate of the fused track follows the track from sensor~2 where sensor~1 does not provide data. In the period where both sensors provide data, the wavrg. provides improved estimates compared to sensor~2's estimate alone.}
\label{fig_t2tResult}
\end{figure}

Fig.~\ref{fig_t2tResult} shows the heading and its variance over time for the walk of the pedestrian shown in Fig.~\ref{fig_t2tPosition}. The sensor generates the heading from the direction of the velocity estimate. The variance of the heading is obtained by linearising the variance of the velocity. Consequently, the variance of the heading is high in the first second for both tracks until the velocity of the track settles. The effect of linearisation is strong, the distribution of the heading will not follow a $\mathcal{VM}$ or $\mathcal{WN}$ distribution as it was our assumption in the previous MC analysis. We do not have access to sensors that yield a proper heading estimation, taking e.g. the object geometry into account, but we expect that future sensors will have such estimations, providing heading that can be properly modelled by a $\mathcal{VM}$ or $\mathcal{WN}$. 
\addtolength{\textheight}{-7.5cm}   

Fig.~\ref{fig_t2tResult} indicates at 1\,s that the crossover problem of the heading gets resolved by the fusion operator. From 1 to 2\,s, the variance is still high, but the estimate for the heading improves significantly due to the fusion.
From 2 to 5\,s, the object is well visible for both sensors, cutting the variance of the fused track in half compared to the track of the single sensors.
From 5\,s onward, sensor~1 needs to predict its track more often. The object is slowly leaving the FoV of the sensor. This leads to an increase in the heading variance by the time the Kalman filter of the single sensor can not update its (predicted) track with new measurement data anymore. At 6.5\,s, the object has left the FoV of sensor~1 completely. The fused heading now relies on the estimate of sensor~2.
The fusion of the heading of the tracks as well as their variances works over the entire period of measurement.

\section{Conclusion}
We discussed multiple measures for the sample variance of circular values before deriving a fusion operator for a high level fusion for both, the von-Mises and the Wrapped Normal distribution.
The fusion operators were analysed using Monte Carlo experiments. We found that the suggested fusion operators are more precise for the Wrapped-Normal than for the von-Mises distribution, since the von-Mises distribution is closed under convolution. The smaller the dispersion and the higher the concentration the better the result of the fusion operators match the Monte Carlo simulation, respectively. 
A track fusion scheme for linear quantities was adapted for circular quantities and tested on real world data. 
The circular fusion operator can work side by side with the operator for linear quantities. 
The experiments showed that our high level fusion operators provide, in contrast to others described in literature, a valid result even after multiple high-level fusions.

\medskip

\bibliographystyle{unsrt}
\bibliography{its_perception}

\vspace{12pt}


\end{document}